\pdfoutput=1
\documentclass[11pt,twoside,a4paper,pdftex,cmspaper,final,collab]{cms-tdr}

\begin{document}\cmsNoteHeader{CFT-09-020}
%
%
%

%
%
\hyphenation{env-iron-men-tal}
\hyphenation{had-ron-i-za-tion}
\hyphenation{cal-or-i-me-ter}
\hyphenation{de-vices}
%
%
\RCS$Revision: 1.25 $
\RCS$Date: 2010/01/11 21:15:34 $
\RCS$Name:  $
%
%
%

\newcommand {\etal}{\mbox{et al.}\xspace} 
\newcommand {\ie}{\mbox{i.e.}\xspace}     
\newcommand {\eg}{\mbox{e.g.}\xspace}     
\newcommand {\etc}{\mbox{etc.}\xspace}     
\newcommand {\vs}{\mbox{\sl vs.}\xspace}      
\newcommand {\mdash}{\ensuremath{\mathrm{-}}} 

\newcommand {\Lone}{Level-1\xspace} 
\newcommand {\Ltwo}{Level-2\xspace}
\newcommand {\Lthree}{Level-3\xspace}

\providecommand{\ACERMC} {\textsc{AcerMC}\xspace}
\providecommand{\ALPGEN} {{\textsc{alpgen}}\xspace}
\providecommand{\CHARYBDIS} {{\textsc{charybdis}}\xspace}
\providecommand{\CMKIN} {\textsc{cmkin}\xspace}
\providecommand{\CMSIM} {{\textsc{cmsim}}\xspace}
\providecommand{\CMSSW} {{\textsc{cmssw}}\xspace}
\providecommand{\COBRA} {{\textsc{cobra}}\xspace}
\providecommand{\COCOA} {{\textsc{cocoa}}\xspace}
\providecommand{\COMPHEP} {\textsc{CompHEP}\xspace}
\providecommand{\EVTGEN} {{\textsc{evtgen}}\xspace}
\providecommand{\FAMOS} {{\textsc{famos}}\xspace}
\providecommand{\GARCON} {\textsc{garcon}\xspace}
\providecommand{\GARFIELD} {{\textsc{garfield}}\xspace}
\providecommand{\GEANE} {{\textsc{geane}}\xspace}
\providecommand{\GEANTfour} {{\textsc{geant4}}\xspace}
\providecommand{\GEANTthree} {{\textsc{geant3}}\xspace}
\providecommand{\GEANT} {{\textsc{geant}}\xspace}
\providecommand{\HDECAY} {\textsc{hdecay}\xspace}
\providecommand{\HERWIG} {{\textsc{herwig}}\xspace}
\providecommand{\HIGLU} {{\textsc{higlu}}\xspace}
\providecommand{\HIJING} {{\textsc{hijing}}\xspace}
\providecommand{\IGUANA} {\textsc{iguana}\xspace}
\providecommand{\ISAJET} {{\textsc{isajet}}\xspace}
\providecommand{\ISAPYTHIA} {{\textsc{isapythia}}\xspace}
\providecommand{\ISASUGRA} {{\textsc{isasugra}}\xspace}
\providecommand{\ISASUSY} {{\textsc{isasusy}}\xspace}
\providecommand{\ISAWIG} {{\textsc{isawig}}\xspace}
\providecommand{\MADGRAPH} {\textsc{MadGraph}\xspace}
\providecommand{\MCATNLO} {\textsc{mc@nlo}\xspace}
\providecommand{\MCFM} {\textsc{mcfm}\xspace}
\providecommand{\MILLEPEDE} {{\textsc{millepede}}\xspace}
\providecommand{\ORCA} {{\textsc{orca}}\xspace}
\providecommand{\OSCAR} {{\textsc{oscar}}\xspace}
\providecommand{\PHOTOS} {\textsc{photos}\xspace}
\providecommand{\PROSPINO} {\textsc{prospino}\xspace}
\providecommand{\PYTHIA} {{\textsc{pythia}}\xspace}
\providecommand{\SHERPA} {{\textsc{sherpa}}\xspace}
\providecommand{\TAUOLA} {\textsc{tauola}\xspace}
\providecommand{\TOPREX} {\textsc{TopReX}\xspace}
\providecommand{\XDAQ} {{\textsc{xdaq}}\xspace}

\newcommand {\DZERO}{D\O\xspace}     


\newcommand{\de}{\ensuremath{^\circ}}
\newcommand{\ten}[1]{\ensuremath{\times \text{10}^\text{#1}}}
\newcommand{\unit}[1]{\ensuremath{\text{\,#1}}\xspace}
\newcommand{\mum}{\ensuremath{\,\mu\text{m}}\xspace}
\newcommand{\micron}{\ensuremath{\,\mu\text{m}}\xspace}
\newcommand{\cm}{\ensuremath{\,\text{cm}}\xspace}
\newcommand{\mm}{\ensuremath{\,\text{mm}}\xspace}
\newcommand{\mus}{\ensuremath{\,\mu\text{s}}\xspace}
\newcommand{\keV}{\ensuremath{\,\text{ke\hspace{-.08em}V}}\xspace}
\newcommand{\MeV}{\ensuremath{\,\text{Me\hspace{-.08em}V}}\xspace}
\newcommand{\GeV}{\ensuremath{\,\text{Ge\hspace{-.08em}V}}\xspace}
\newcommand{\TeV}{\ensuremath{\,\text{Te\hspace{-.08em}V}}\xspace}
\newcommand{\PeV}{\ensuremath{\,\text{Pe\hspace{-.08em}V}}\xspace}
\newcommand{\keVc}{\ensuremath{{\,\text{ke\hspace{-.08em}V\hspace{-0.16em}/\hspace{-0.08em}c}}}\xspace}
\newcommand{\MeVc}{\ensuremath{{\,\text{Me\hspace{-.08em}V\hspace{-0.16em}/\hspace{-0.08em}c}}}\xspace}
\newcommand{\GeVc}{\ensuremath{{\,\text{Ge\hspace{-.08em}V\hspace{-0.16em}/\hspace{-0.08em}c}}}\xspace}
\newcommand{\TeVc}{\ensuremath{{\,\text{Te\hspace{-.08em}V\hspace{-0.16em}/\hspace{-0.08em}c}}}\xspace}
\newcommand{\keVcc}{\ensuremath{{\,\text{ke\hspace{-.08em}V\hspace{-0.16em}/\hspace{-0.08em}c}^\text{2}}}\xspace}
\newcommand{\MeVcc}{\ensuremath{{\,\text{Me\hspace{-.08em}V\hspace{-0.16em}/\hspace{-0.08em}c}^\text{2}}}\xspace}
\newcommand{\GeVcc}{\ensuremath{{\,\text{Ge\hspace{-.08em}V\hspace{-0.16em}/\hspace{-0.08em}c}^\text{2}}}\xspace}
\newcommand{\TeVcc}{\ensuremath{{\,\text{Te\hspace{-.08em}V\hspace{-0.16em}/\hspace{-0.08em}c}^\text{2}}}\xspace}

\newcommand{\pbinv} {\mbox{\ensuremath{\,\text{pb}^\text{$-$1}}}\xspace}
\newcommand{\fbinv} {\mbox{\ensuremath{\,\text{fb}^\text{$-$1}}}\xspace}
\newcommand{\nbinv} {\mbox{\ensuremath{\,\text{nb}^\text{$-$1}}}\xspace}
\newcommand{\percms}{\ensuremath{\,\text{cm}^\text{$-$2}\,\text{s}^\text{$-$1}}\xspace}
\newcommand{\lumi}{\ensuremath{\mathcal{L}}\xspace}
\newcommand{\Lumi}{\ensuremath{\mathcal{L}}\xspace}
%
\newcommand{\LvLow}  {\ensuremath{\mathcal{L}=\text{10}^\text{32}\,\text{cm}^\text{$-$2}\,\text{s}^\text{$-$1}}\xspace}
\newcommand{\LLow}   {\ensuremath{\mathcal{L}=\text{10}^\text{33}\,\text{cm}^\text{$-$2}\,\text{s}^\text{$-$1}}\xspace}
\newcommand{\lowlumi}{\ensuremath{\mathcal{L}=\text{2}\times \text{10}^\text{33}\,\text{cm}^\text{$-$2}\,\text{s}^\text{$-$1}}\xspace}
\newcommand{\LMed}   {\ensuremath{\mathcal{L}=\text{2}\times \text{10}^\text{33}\,\text{cm}^\text{$-$2}\,\text{s}^\text{$-$1}}\xspace}
\newcommand{\LHigh}  {\ensuremath{\mathcal{L}=\text{10}^\text{34}\,\text{cm}^\text{$-$2}\,\text{s}^\text{$-$1}}\xspace}
\newcommand{\hilumi} {\ensuremath{\mathcal{L}=\text{10}^\text{34}\,\text{cm}^\text{$-$2}\,\text{s}^\text{$-$1}}\xspace}


\newcommand{\zp}{\ensuremath{\mathrm{Z}^\prime}\xspace}


\newcommand{\kt}{\ensuremath{k_{\mathrm{T}}}\xspace}
\newcommand{\BC}{\ensuremath{{B_{\mathrm{c}}}}\xspace}
\newcommand{\bbarc}{\ensuremath{{\overline{b}c}}\xspace}
\newcommand{\bbbar}{\ensuremath{{b\overline{b}}}\xspace}
\newcommand{\ccbar}{\ensuremath{{c\overline{c}}}\xspace}
\newcommand{\JPsi}{\ensuremath{{J}/\psi}\xspace}
\newcommand{\bspsiphi}{\ensuremath{B_s \to \JPsi\, \phi}\xspace}
\newcommand{\AFB}{\ensuremath{A_\mathrm{FB}}\xspace}
\newcommand{\EE}{\ensuremath{e^+e^-}\xspace}
\newcommand{\MM}{\ensuremath{\mu^+\mu^-}\xspace}
\newcommand{\TT}{\ensuremath{\tau^+\tau^-}\xspace}
\newcommand{\wangle}{\ensuremath{\sin^{2}\theta_{\mathrm{eff}}^\mathrm{lept}(M^2_\mathrm{Z})}\xspace}
\newcommand{\ttbar}{\ensuremath{{t\overline{t}}}\xspace}
\newcommand{\stat}{\ensuremath{\,\text{(stat.)}}\xspace}
\newcommand{\syst}{\ensuremath{\,\text{(syst.)}}\xspace}

\newcommand{\HGG}{\ensuremath{\mathrm{H}\to\gamma\gamma}}
\newcommand{\gev}{\GeV}
\newcommand{\GAMJET}{\ensuremath{\gamma + \mathrm{jet}}}
\newcommand{\PPTOJETS}{\ensuremath{\mathrm{pp}\to\mathrm{jets}}}
\newcommand{\PPTOGG}{\ensuremath{\mathrm{pp}\to\gamma\gamma}}
\newcommand{\PPTOGAMJET}{\ensuremath{\mathrm{pp}\to\gamma +
\mathrm{jet}
}}
\newcommand{\MH}{\ensuremath{\mathrm{M_{\mathrm{H}}}}}
\newcommand{\RNINE}{\ensuremath{\mathrm{R}_\mathrm{9}}}
\newcommand{\DR}{\ensuremath{\Delta\mathrm{R}}}


\newcommand{\PT}{\ensuremath{p_{\mathrm{T}}}\xspace}
\newcommand{\pt}{\ensuremath{p_{\mathrm{T}}}\xspace}
\newcommand{\ET}{\ensuremath{E_{\mathrm{T}}}\xspace}
\newcommand{\HT}{\ensuremath{H_{\mathrm{T}}}\xspace}
\newcommand{\et}{\ensuremath{E_{\mathrm{T}}}\xspace}
\newcommand{\Em}{\ensuremath{E\!\!\!/}\xspace}
\newcommand{\Pm}{\ensuremath{p\!\!\!/}\xspace}
\newcommand{\PTm}{\ensuremath{{p\!\!\!/}_{\mathrm{T}}}\xspace}
\newcommand{\ETm}{\ensuremath{E_{\mathrm{T}}^{\mathrm{miss}}}\xspace}
\newcommand{\MET}{\ensuremath{E_{\mathrm{T}}^{\mathrm{miss}}}\xspace}
\newcommand{\ETmiss}{\ensuremath{E_{\mathrm{T}}^{\mathrm{miss}}}\xspace}
\newcommand{\VEtmiss}{\ensuremath{{\vec E}_{\mathrm{T}}^{\mathrm{miss}}}\xspace}

%

\newcommand{\ga}{\ensuremath{\gtrsim}}
\newcommand{\la}{\ensuremath{\lesssim}}
\newcommand{\swsq}{\ensuremath{\sin^2\theta_W}\xspace}
\newcommand{\cwsq}{\ensuremath{\cos^2\theta_W}\xspace}
\newcommand{\tanb}{\ensuremath{\tan\beta}\xspace}
\newcommand{\tanbsq}{\ensuremath{\tan^{2}\beta}\xspace}
\newcommand{\sidb}{\ensuremath{\sin 2\beta}\xspace}
\newcommand{\alpS}{\ensuremath{\alpha_S}\xspace}
\newcommand{\alpt}{\ensuremath{\tilde{\alpha}}\xspace}

\newcommand{\QL}{\ensuremath{Q_L}\xspace}
\newcommand{\sQ}{\ensuremath{\tilde{Q}}\xspace}
\newcommand{\sQL}{\ensuremath{\tilde{Q}_L}\xspace}
\newcommand{\ULC}{\ensuremath{U_L^C}\xspace}
\newcommand{\sUC}{\ensuremath{\tilde{U}^C}\xspace}
\newcommand{\sULC}{\ensuremath{\tilde{U}_L^C}\xspace}
\newcommand{\DLC}{\ensuremath{D_L^C}\xspace}
\newcommand{\sDC}{\ensuremath{\tilde{D}^C}\xspace}
\newcommand{\sDLC}{\ensuremath{\tilde{D}_L^C}\xspace}
\newcommand{\LL}{\ensuremath{L_L}\xspace}
\newcommand{\sL}{\ensuremath{\tilde{L}}\xspace}
\newcommand{\sLL}{\ensuremath{\tilde{L}_L}\xspace}
\newcommand{\ELC}{\ensuremath{E_L^C}\xspace}
\newcommand{\sEC}{\ensuremath{\tilde{E}^C}\xspace}
\newcommand{\sELC}{\ensuremath{\tilde{E}_L^C}\xspace}
\newcommand{\sEL}{\ensuremath{\tilde{E}_L}\xspace}
\newcommand{\sER}{\ensuremath{\tilde{E}_R}\xspace}
\newcommand{\sFer}{\ensuremath{\tilde{f}}\xspace}
\newcommand{\sQua}{\ensuremath{\tilde{q}}\xspace}
\newcommand{\sUp}{\ensuremath{\tilde{u}}\xspace}
\newcommand{\suL}{\ensuremath{\tilde{u}_L}\xspace}
\newcommand{\suR}{\ensuremath{\tilde{u}_R}\xspace}
\newcommand{\sDw}{\ensuremath{\tilde{d}}\xspace}
\newcommand{\sdL}{\ensuremath{\tilde{d}_L}\xspace}
\newcommand{\sdR}{\ensuremath{\tilde{d}_R}\xspace}
\newcommand{\sTop}{\ensuremath{\tilde{t}}\xspace}
\newcommand{\stL}{\ensuremath{\tilde{t}_L}\xspace}
\newcommand{\stR}{\ensuremath{\tilde{t}_R}\xspace}
\newcommand{\stone}{\ensuremath{\tilde{t}_1}\xspace}
\newcommand{\sttwo}{\ensuremath{\tilde{t}_2}\xspace}
\newcommand{\sBot}{\ensuremath{\tilde{b}}\xspace}
\newcommand{\sbL}{\ensuremath{\tilde{b}_L}\xspace}
\newcommand{\sbR}{\ensuremath{\tilde{b}_R}\xspace}
\newcommand{\sbone}{\ensuremath{\tilde{b}_1}\xspace}
\newcommand{\sbtwo}{\ensuremath{\tilde{b}_2}\xspace}
\newcommand{\sLep}{\ensuremath{\tilde{l}}\xspace}
\newcommand{\sLepC}{\ensuremath{\tilde{l}^C}\xspace}
\newcommand{\sEl}{\ensuremath{\tilde{e}}\xspace}
\newcommand{\sElC}{\ensuremath{\tilde{e}^C}\xspace}
\newcommand{\seL}{\ensuremath{\tilde{e}_L}\xspace}
\newcommand{\seR}{\ensuremath{\tilde{e}_R}\xspace}
\newcommand{\snL}{\ensuremath{\tilde{\nu}_L}\xspace}
\newcommand{\sMu}{\ensuremath{\tilde{\mu}}\xspace}
\newcommand{\sNu}{\ensuremath{\tilde{\nu}}\xspace}
\newcommand{\sTau}{\ensuremath{\tilde{\tau}}\xspace}
\newcommand{\Glu}{\ensuremath{g}\xspace}
\newcommand{\sGlu}{\ensuremath{\tilde{g}}\xspace}
\newcommand{\Wpm}{\ensuremath{W^{\pm}}\xspace}
\newcommand{\sWpm}{\ensuremath{\tilde{W}^{\pm}}\xspace}
\newcommand{\Wz}{\ensuremath{W^{0}}\xspace}
\newcommand{\sWz}{\ensuremath{\tilde{W}^{0}}\xspace}
\newcommand{\sWino}{\ensuremath{\tilde{W}}\xspace}
\newcommand{\Bz}{\ensuremath{B^{0}}\xspace}
\newcommand{\sBz}{\ensuremath{\tilde{B}^{0}}\xspace}
\newcommand{\sBino}{\ensuremath{\tilde{B}}\xspace}
\newcommand{\Zz}{\ensuremath{Z^{0}}\xspace}
\newcommand{\sZino}{\ensuremath{\tilde{Z}^{0}}\xspace}
\newcommand{\sGam}{\ensuremath{\tilde{\gamma}}\xspace}
\newcommand{\chiz}{\ensuremath{\tilde{\chi}^{0}}\xspace}
\newcommand{\chip}{\ensuremath{\tilde{\chi}^{+}}\xspace}
\newcommand{\chim}{\ensuremath{\tilde{\chi}^{-}}\xspace}
\newcommand{\chipm}{\ensuremath{\tilde{\chi}^{\pm}}\xspace}
\newcommand{\Hone}{\ensuremath{H_{d}}\xspace}
\newcommand{\sHone}{\ensuremath{\tilde{H}_{d}}\xspace}
\newcommand{\Htwo}{\ensuremath{H_{u}}\xspace}
\newcommand{\sHtwo}{\ensuremath{\tilde{H}_{u}}\xspace}
\newcommand{\sHig}{\ensuremath{\tilde{H}}\xspace}
\newcommand{\sHa}{\ensuremath{\tilde{H}_{a}}\xspace}
\newcommand{\sHb}{\ensuremath{\tilde{H}_{b}}\xspace}
\newcommand{\sHpm}{\ensuremath{\tilde{H}^{\pm}}\xspace}
\newcommand{\hz}{\ensuremath{h^{0}}\xspace}
\newcommand{\Hz}{\ensuremath{H^{0}}\xspace}
\newcommand{\Az}{\ensuremath{A^{0}}\xspace}
\newcommand{\Hpm}{\ensuremath{H^{\pm}}\xspace}
\newcommand{\sGra}{\ensuremath{\tilde{G}}\xspace}
\newcommand{\mtil}{\ensuremath{\tilde{m}}\xspace}
\newcommand{\rpv}{\ensuremath{\rlap{\kern.2em/}R}\xspace}
\newcommand{\LLE}{\ensuremath{LL\bar{E}}\xspace}
\newcommand{\LQD}{\ensuremath{LQ\bar{D}}\xspace}
\newcommand{\UDD}{\ensuremath{\overline{UDD}}\xspace}
\newcommand{\Lam}{\ensuremath{\lambda}\xspace}
\newcommand{\Lamp}{\ensuremath{\lambda'}\xspace}
\newcommand{\Lampp}{\ensuremath{\lambda''}\xspace}
\newcommand{\spinbd}[2]{\ensuremath{\bar{#1}_{\dot{#2}}}\xspace}

\newcommand{\MD}{\ensuremath{{M_\mathrm{D}}}\xspace}
\newcommand{\Mpl}{\ensuremath{{M_\mathrm{Pl}}}\xspace}
\newcommand{\Rinv} {\ensuremath{{R}^{-1}}\xspace}

%
%
\hyphenation{en-viron-men-tal}

\cmsNoteHeader{CFT-09-020}
\title{Commissioning of the CMS High-Level Trigger with Cosmic Rays}


\date{\today}

\abstract{
   The CMS High-Level Trigger (HLT) is responsible for ensuring that data samples with potentially interesting events are recorded with high efficiency and good quality. This paper gives an overview of the HLT and focuses on its commissioning using cosmic rays. The selection of triggers that were deployed is presented and the online grouping of triggered events into streams and primary datasets is discussed. Tools for online and offline data quality monitoring for the HLT are described, and  the operational performance of the muon HLT algorithms is reviewed. The average time taken for the HLT selection and its dependence on detector and operating conditions are presented. The HLT performed reliably and helped provide a large dataset. This dataset has proven to be invaluable for understanding the performance of the trigger and the CMS experiment as a whole.
}

\hypersetup{%
pdfauthor={Trigger Studies Group},%
pdftitle={Commissioning of the CMS High-Level Trigger with Cosmic Rays},%
pdfsubject={CMS},%
pdfkeywords={CMS, HLT}}

\maketitle 

\section{Introduction}
\label{sec:intro}

The Compact Muon Solenoid (CMS) detector~\cite{JINST} is one of two large general-purpose experiments at the CERN Large Hadron Collider (LHC)~\cite{Evans:2008zzb}. The 12,500~ton detector features nearly complete solid-angle coverage and can precisely measure electrons, photons, muons, jets and missing energy over a large range of particle energies. These broad capabilities of the CMS detector will allow the exploration of electroweak symmetry breaking and enable the potential discovery of physics beyond the Standard Model~\cite{ptdr}.

The CMS trigger and data acquisition systems are responsible for ensuring that data samples with potentially interesting events are recorded with high efficiency and good quality. The experiment has a two-level trigger system, unlike most other hadron collider experiments which have more traditional three-level systems. The first level is hardware-based and is called the ``Level-1 Trigger'' (L1) while the second  level is software-based and is called the ``High-Level Trigger'' (HLT).

In late 2008 the CMS Collaboration conducted a month-long data taking exercise known as the Cosmic Run At Four Tesla (CRAFT) with the goal of  commissioning the experiment for extended operation~\cite{Overview}. With all installed detector systems participating, CMS recorded 270 million cosmic ray triggered events with the solenoid at its nominal axial field strength of 3.8~T. In addition to commissioning the experiment operationally, the data collected during CRAFT proved invaluable for understanding the performance of the CMS experiment as a whole.

This paper reviews the implementation of the CMS HLT and its commissioning using CRAFT data. For the first time, during CRAFT, the CMS HLT was run over an extended period of time utilizing data from most sub-detectors. In addition to the gain in operational experience, the data collected  proved to be extremely useful for developing monitoring strategies and procedures. In particular, since the physics data were primarily comprised of muons, muon-based HLT algorithms benefited most.

The paper is organized as follows. The CMS Trigger system is described in Section~\ref{sec:trigger} and an overview of the triggers deployed online during cosmic ray data taking is provided in Section~\ref{sec:menus}. The output of the HLT is reviewed in Section~\ref{sec:StreamsPDs} while the strategies implemented and commissioned for monitoring the performance of the HLT are discussed in detail in Section~\ref{sec:performance}.

\section{The CMS Detector and its Trigger System}
\label{sec:trigger}
The central feature of the Compact Muon Solenoid (CMS) apparatus is a superconducting solenoid, of 6~m internal diameter, providing a field of 3.8~T. Within the field volume are the silicon pixel and strip tracker, the crystal electromagnetic calorimeter (ECAL) and the brass/scintillator hadron calorimeter (HCAL). Muons are measured in the pseudorapidity window $|\eta|< 2.4$, with detection planes made of three technologies: Drift Tubes (DT), Cathode Strip Chambers (CSC), and Resistive Plate Chambers (RPC). In addition to the barrel and endcap detectors, CMS has extensive forward calorimetry (HF).

\begin{figure}[!htbp]
   \begin{center}
      \includegraphics[width=0.5\linewidth]{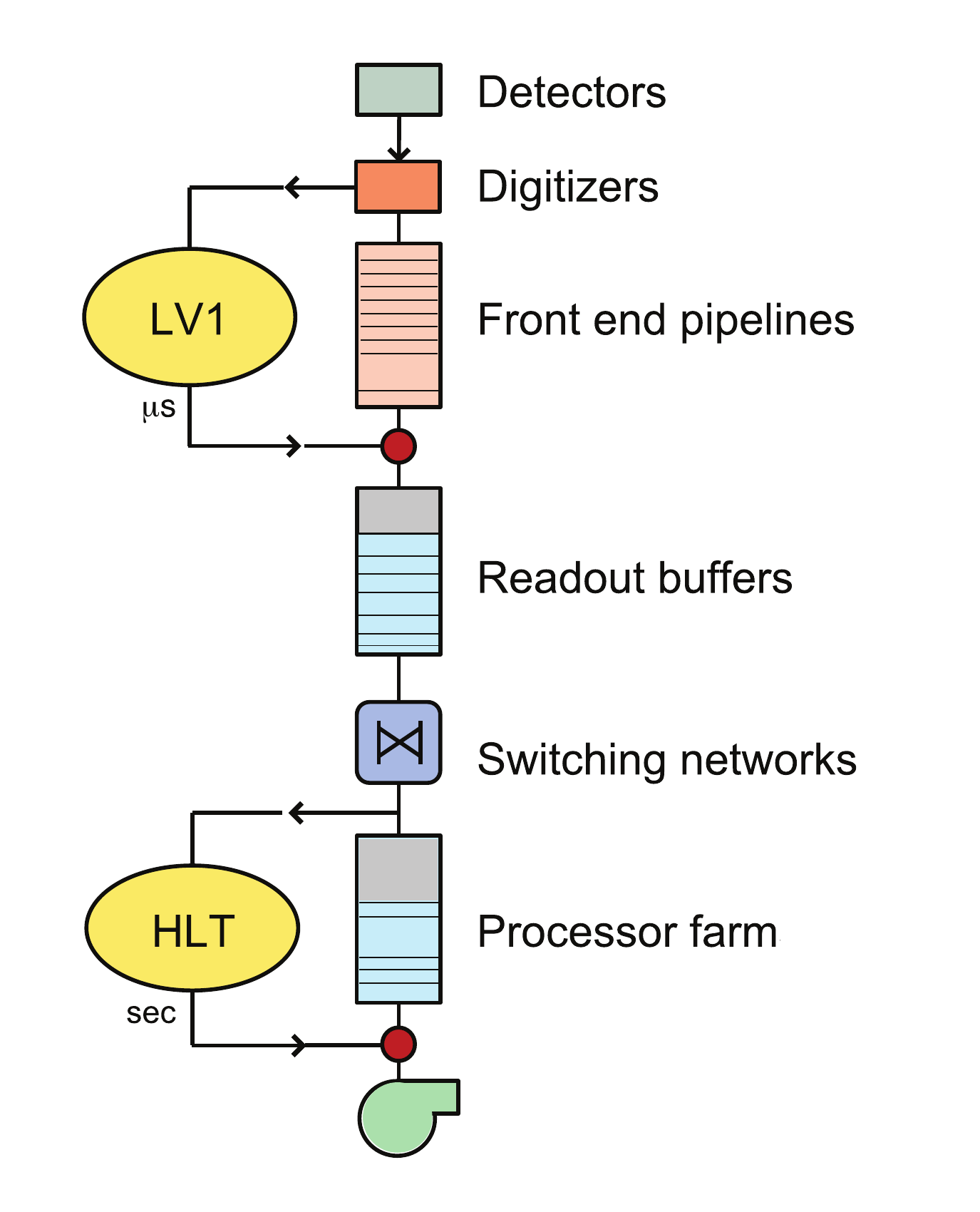}
      \caption{Schematic view of the CMS trigger system.}
   \label{fig:trigger}
   \end{center}
\end{figure}

A schematic view of the CMS trigger system is presented in Fig.~\ref{fig:trigger}. The details of the trigger design and its implementation can be found in Refs.~\cite{L1tdr,tdr,CMS_AN_2007-021}. L1 is designed to process information from the CMS detector at a bunch-crossing rate of 40~MHz. This rate, coupled with the latency of 3\mus, limits the data accessible for processing to the information from the calorimeter and muon detectors. Additionally, in order to allow fast processing, these data have coarser granularity and lower resolution than the full information from the front-end electronics. The L1 trigger is built mostly of custom hardware and is designed to reduce the event rate to 100~kHz at the LHC design luminosity \LHigh.

The CMS HLT hardware consists of a single processor farm called the ``Event Filter Farm''. The farm comprises PCs mounting 2.66~GHz dual quad-core Intel processors and a total of 16 GB RAM. The HLT algorithms that run on the  farm are required to reduce the event rate to $O(100)$~Hz. In traditional three-level trigger systems this rate reduction of $O(1000)$ is achieved in two stages: a customized hardware or software-based  Level-2 trigger that quickly provides the large rejection factor and a Level-3 processor farm that makes decisions based on more sophisticated algorithms.  The CMS HLT, in contrast, combines the rejection power and speed of a Level-2 trigger with the flexibility and sophistication of Level-3 in a single processor farm. The design of the filter farm is described in detail in Ref.~\cite{EVFcommission}.

Traditional three-level systems have to use ``regional'' information from specific parts of the detector to achieve the $O(1000)$ rejection. Instead, in the CMS  two-level design, full-granularity data from the whole detector (including the tracker) are available to a standard commercial computer. Regional data-unpacking is implemented only for some of the sub-detectors to optimize the CPU performance of the HLT.  The CMS HLT decision can then be based on full resolution data, and the filtering algorithms have the flexibility to use sophisticated offline-quality reconstruction algorithms. Due to the high input rate of 100~kHz which the filter farm needs to sustain, this implementation requires significant CPU resources.  Therefore, the reconstruction algorithms that run during the HLT processing must be optimized for performance in order to minimize dead-time and, at the same time, collect all interesting events for physics analysis.

The data acquisition (DAQ) system commissioned during CRAFT consisted of a total of 275
builder units that assembled the data fragments into complete events, and 825 filter units for the HLT processing. Towards the end of CRAFT, a DAQ configuration of 720 8-core PCs and nearly 4500 filter units was successfully tested. This DAQ configuration is planned to be used at LHC startup. With such a configuration, the CPU power available in the filter farm is expected to be suitable for handling HLT input rates as high as 50~kHz for an average processing time of up to $\approx$100~ms/event. This estimate includes a conservative factor of two in the average HLT processing time which has been measured to be  $\approx$43~ms using fully simulated events~\cite{CMS_AN_2007-021}. For LHC operation at high luminosity, the computing power of the filter farm is expected to double in size, with the use of more recent CPUs, thereby meeting the design HLT input rate of $\sim 100$~kHz.



\section{HLT during CRAFT}
\label{sec:menus}
Candidate HLT ``menus'' for startup were commissioned during CRAFT and their performance was studied. A menu is defined as a set of triggers, each of which consists of several reconstruction and filtering modules. An event is stored to permanent storage if it is selected by at least one High-Level trigger in the deployed  menus.

The HLT menus deployed during CRAFT were designed to reliably deliver data from cosmic rays, regardless of detector
conditions.
Consequently,
these menus were primarily composed of ``L1 pass-through''
triggers, which transfer to storage the events accepted by one or more L1
triggers without additional processing or selection. Triggers with additional
selection beyond the L1 decision were added to the HLT menus throughout the commissioning
period.   In subsequent studies of the recorded data, these triggers
were useful for
physics selection, detector diagnostics, and the alignment and calibration of
the CMS subsystems. 

Prior to CRAFT, HLT algorithms had been studied using Monte Carlo simulated
data.  The introduction of these algorithms to the HLT menus during CRAFT
provided an opportunity to test their behavior in the detector environment and
improve the selection criteria.

\subsection{CRAFT triggers}
\label{sec:collisionmenu}

Every event processed by the HLT must first satisfy one or more of the
L1 trigger requirements.  The L1 Global Trigger (GT)~\cite{CRAFT_L1} can
provide up to 128  trigger algorithms that select an event, based on logical
combinations of L1 trigger objects (such as muons, jets, or calorimeter
transverse energy) with applied selection criteria (energy/momentum
thresholds, etc.).  Up to 64 simple on/off signals from the sub-detectors, or
``Technical triggers'', can also be added to the GT decision. The
technical triggers provide additional information that may be used
for special
purposes such as detector diagnostics and monitoring.
During CRAFT, only the most inclusive L1 single-object trigger algorithms contributed to the global L1 decision: a muon (with no
additional selection), a jet with transverse energy above 10~GeV, or an
electron or photon with at least 1~GeV of transverse energy.  The results of
the remaining L1 trigger algorithms were computed but were not considered in the
GT decision, i.e., masked. The full L1 results (including these masked triggers) were available to the
HLT, thus allowing the more complex HLT
triggers to be seeded by more restrictive L1 trigger algorithms.

The HLT menus deployed during CRAFT included triggers that can be grouped into two
main categories: L1 pass-through triggers, and triggers that required some
additional processing beyond the L1 decision.  All HLT triggers were run over all events accepted at L1.

For most L1 pass-through triggers, the HLT only checked the event
type as marked by the GT.  The HLT accepted events if any of the following L1
conditions was satisfied:
\begin{itemize}
  \item The event was marked as a ``physics'' event by the GT.  This required a
  positive response from any of the trigger algorithms or technical triggers that
  contributed to the GT decision. This ensured that all events accepted at L1 were kept for later
  study. These events occurred at a rate of $\sim 500$~Hz.
  \item The event was marked as a ``calibration'' event.  These
  specially triggered events occur at the end of about 1\% of all LHC orbits during the  3\mus gap that is devoid of protons in both LHC beams~\cite{LHCbunch}.
  This orbit gap was emulated by CMS during CRAFT.  All L1 ``physics'' triggers (algorithms
  or technical triggers) are suppressed during a calibration event.  These
  events occur at a rate of 100~Hz and are used for detector
  calibration and diagnostics.
  \item The event was marked by the GT as a ``random'' trigger event.  These
  events are sent to the HLT in order to verify the L1 decision.  The random
  trigger rate was limited to a few Hz by the HLT.
\end{itemize}

Three HLT pass-through triggers were implemented, one for each of the above conditions. The other HLT triggers included in the CRAFT menus selected
sub-samples of the data already accepted by these triggers.  This allowed the data to
be organized into
more manageable blocks (see Section~\ref{sec:StreamsPDs}).

In addition to the L1 pass-through triggers described above, several more
complicated triggers
were included in the HLT menus to help with detector diagnostics:

\begin{itemize}
  \item
  A trigger, seeded by random L1 triggers, accepted events
  based on the size of the pixel tracker data.  This sub-detector has a very low
  noise rate, such that charged particles passing
  through the tracker noticeably increase the pixel data
  volume.  This trigger could be used to validate the L1 decision, since such an event, with particles crossing the pixel
  tracker, could be accepted by the HLT even if it was not flagged as ``physics'' by the GT.
  \item Similarly, a trigger was designed to accept events with energy deposition in HF exceeding the noise threshold.
  \item During dedicated runs designed to monitor HCAL noise behavior, the L1
  trigger system was configured to accept events consistent with noise from
  HCAL photodiodes.  HLT CRAFT menus included a pass-through trigger to
  accept these events.
\end{itemize}

In order
to commission the HLT reconstruction algorithms in preparation for collisions,
physics triggers that applied additional selection criteria, beyond the L1 seed
requirement, were also included in the HLT menu.  These triggers include
selection algorithms that will be used to filter collision data, but with relaxed criteria (energy/momentum thresholds, etc.) in order to collect
cosmic ray and detector noise data.

\begin{itemize}
  \item For events accepted by the GT with a L1 jet algorithm, the calorimeter
  data in the vicinity of the L1 jet were used to reconstruct jet candidates
  in the HLT using the CMS implementation of the Iterative Cone algorithm~\cite{JetICone}.  The HLT
  jet transverse energy was required to exceed 12~GeV.
  \item Events that triggered the L1 jet algorithms were accepted by the HLT
  without additional selection criteria.  These events, collected without
  beam passing through CMS, will be used for
  detector background studies, relevant for example for heavy stable charged particle searches
  in collision data~\cite{gluino}.
  \item For events with an electron/photon L1 seed, the ECAL regions
  surrounding the L1 seed were examined and an electromagnetic object was reconstructed using calorimeter information only.  The event was
  accepted by the HLT if the electron/photon HLT object had at least
  5~GeV of transverse energy.  Information from the tracker was ignored.
  \item Several muon triggers were included in the CRAFT menus:
  \begin{itemize}
    \item[$\star$] The muon detector data were examined
    and the muon was reconstructed from the L1 muon seed using this information.  Muons with
    at least 3\GeVc of transverse momentum ($p_{T}$) were accepted.
    \item[$\star$] Starting with the L1 seed in the muon detector, the path of the
  reconstructed muon was  extrapolated inward to the tracker volume.
  Information from the reconstructed particle trajectory within the tracker 
    was used to improve the estimation of the muon momentum.  Events with muons with $p_{T} > 3$\GeVc were
    accepted.
    \item[$\star$] Alternatively, tracks were reconstructed in the tracker
    volume
    without seeding from a L1 muon object.  Events accepted by this trigger,
  which examined any event accepted by L1,
    provided a source of cosmic rays intersecting the tracker volume that may
    have been missed by the standard HLT muon selection and can be used
    to improve the HLT reconstruction algorithms.  This trigger could only
    be implemented during commissioning when the HLT input rate is low
    compared to the pp collision rate.  In order to process the large number
    of events expected from L1 during colliding beam operations, any trigger
    that implements some non-trivial selection criteria in the HLT must first
    refine the input using L1 seed information.
  \end{itemize}
\end{itemize}

Finally, triggers were included in the HLT menus to satisfy the alignment
and calibration needs of the CMS experiment.  The performance of these
triggers will be discussed in Section~5.

\subsection{Online deployment of the HLT menu}
\label{sec:Deployment}

The menu that is deployed {\it online} for data taking operations closely
resembles the {\it offline}-tested menu.  The HLT uses the same software to
run offline on Monte Carlo simulations and previously collected  events
and to record online collision or cosmic ray data.  However,
a small set of technical adjustments is necessary when the menu is moved from the
offline computing environment to the Event Filter Farm. These adjustments can potentially introduce a mismatch
between what is tested offline and what is used  online.

The HLT menu is tested on a dedicated section of the Event Filter Farm prior to
online deployment.  This validation
farm emulates the behavior of the full Event Filter Farm and is used to validate the HLT menu
on previously collected data samples.  The
validation farm comprises a Storage Manager (SM) application and an adjustable
number of filter units, up to 19, each running up to four independent event
processing threads. The SM is the last component in the data-handling chain and
performs two primary functions. The first function is to collect events from each
Filter Farm HLT process, and store them in files for subsequent transfer and
processing. The data files are assigned to one or more output streams (see Section~4) according to the defined trigger bits and the
SM configuration
information. The second function of the SM is to act as an event
server for calibration and monitoring purposes.

A sample of previously collected data events is examined using the validation farm,  the events being filtered by the trigger menu under consideration. This procedure validates several aspects of the trigger menu
which cannot be otherwise tested offline:
\begin{itemize}
  \item the trigger menu behavior, when first deployed on the Event Filter
  Farm.  Due to the differences in the online and offline environments, it is
  possible that an offline-validated menu is incompatible with the online
  environment;
  \item the selection of events by the menu, and whether rates are
  consistent with expectations from offline tests;
  \item the  correct grouping of data into streams, each containing the
  specified set of ``primary datasets'' (see Section~4);
  \item the communication with the Storage Manager and the production of the
  appropriate output file for each stream.
\end{itemize}

The online test also provides a useful opportunity for
debugging the candidate HLT menu. The Event Filter Farm places problematic events, such
as events with corrupt data or reconstruction problems resulting in the
failure of the HLT filtering process, into a special ``Error''
stream.  The content of this stream is analyzed offline in order to diagnose problems and to serve as a test for new trigger algorithms.

During CRAFT operation, a menu was first validated with data on the validation farm and then deployed online.  The performance of the menu was then monitored as discussed in
Section~\ref{sec:OnlineDQM}.

A total of 14 HLT menus were deployed during the CRAFT data taking period.  In most
cases, new menus were created by making small adjustments to the existing
menu.  These changes were prompted by the daily operational goals
and could be validated and implemented within an hour.  By contrast, the HLT
menus that included triggers designed to test HLT reconstruction algorithms
were validated over several days to ensure that they could be deployed online
without problems.


\section{Streams and Primary Datasets}
\label{sec:StreamsPDs}

As described in Section~\ref{sec:Deployment}, the SM is responsible for collecting events from each Filter Farm HLT process and storing them in files for subsequent transfer and processing. Events are grouped into a set of streams, with distinct sets of data files produced for each stream. The grouping of events is done according to their offline usage. As an example, the primary stream is ``Stream A'', or the ``physics'' stream, which consists of events which satisfy the needs of most offline physics analyses. Other streams serve special purposes such as monitoring the performance of the HLT algorithms.

Different streams must be defined for samples that require special or reduced amount of information (``event content''). This is necessary since the software framework requires both online and offline data files to contain homogeneous information. For example, the event content used for the physics stream comprises the full detector and trigger raw data, and the  L1 and HLT trigger results. However, streams used for calibration can benefit from storing only a fraction of the raw data, while using triggers with higher accept rates than those included in the physics stream. This allows the highest possible collection rate for calibration events while respecting bandwidth constraints.

The streams that were defined in the HLT configuration and exercised during CRAFT are listed below:

\begin{itemize}
\item Stream A:
The primary physics stream, which consists of detector and trigger  raw data and L1 and HLT trigger summary results. The full list of physics triggers feeds this stream.
\item DQM:
Used to monitor the performance of the detector, the L1 trigger and the HLT algorithms. All relevant by-products of the HLT reconstruction  are stored in order to allow detailed online and offline validation. These reconstruction by-products are objects created by the HLT processing, contain detailed information regarding the execution of the algorithms, and can double the size of the data stream for certain types of events. These by-products are not stored in the Stream A output.
\item Calibration:
Used to collect ``calibration'' type L1-triggered events during the orbit gaps (see Section~\ref{sec:collisionmenu}). The calibration stream provides events for monitoring and calibrating the various sub-detectors.
\item AlCaRaw:
A group of streams used to collect data for calibrating the ECAL and HCAL detectors. Typically, these streams are fed by high rate triggers which have reduced event content. During collisions, these data will be used for providing new calibration and alignment constants for use by offline reconstruction algorithms.The AlCaRaw streams are described in more detail in Section~\ref{sec:AlCaraw}.
\end{itemize}

During CRAFT, a total of seven streams (Stream A, DQM, Calibration, and 4 AlCaRaw streams) were produced. Upon transfer to the CERN ``Tier-0'' computing site, the data corresponding to the various streams was reformatted and split into various ``Primary Datasets''. These datasets are defined, within a stream, by grouping a set of triggers that perform similar selections (e.g., all muon-based triggers within Stream A were grouped into the ``Cosmics'' dataset). The primary datasets are intended to provide small-sized samples suitable for specific physics analyses.

The definition of multiple streams and primary datasets during CRAFT and their subsequent offline processing helped in the commissioning of the full workflow that will be used at LHC startup.

\subsection{DQM stream }

\label{Sec:hltdebug}

To perform  data quality monitoring (DQM), a special stream was used during CRAFT. This stream contained the same triggers  as the
primary physics stream, Stream A, with two main differences.  The DQM stream
included events selected by the calibration triggers, which were part of the calibration
stream.  This allowed the DQM team to check and monitor the quality
of the data used for calibration purposes. In addition, the stream included
the HLT reconstruction by-products. These by-products are essential for monitoring the proper performance of the HLT algorithms  and can be effectively used for debugging purposes.

The DQM stream
is required to provide sufficient data for online DQM (see Section~\ref{sec:OnlineDQM}). The
rate of the stream is determined mainly by the time taken to reformat the data
before it is fed to the DQM applications.
 During CRAFT the rate of the stream was 25~Hz, but for the final operation during collision data taking, a rate of about 60~Hz or higher is expected.

\subsection{AlCaRaw streams}
\label{sec:AlCaraw}

The term ``AlCaRaw'' is used to designate high-rate triggers that provide events for detector alignment and calibration. The name consists of two parts: ``AlCa''
indicates that the events are used for alignment and calibration, and
``Raw'' indicates that the events are selected directly at the raw
data level.

Alignment and calibration techniques typically require large size
samples. However, only a small portion of the event data is used,
i.e., the full event is not needed for most alignment and
calibration purposes. Usually, minimum bias events serve this purpose
best. These have high rates and hence directly compete with the
triggers needed for physics analyses. In order to collect the
statistics required for alignment and calibration, without saturating
the data taking bandwidth, dedicated triggers record events and carry
out data reduction so as to store the minimal amount of information
needed for the calibration or alignment processes. Data collected by
these triggers are used for the calibration of the ECAL and HCAL
sub-detectors.

Four AlCaRaw streams were exercised during CRAFT:
\begin{itemize}
\item ECAL $\pi^0$: uses a constraint on the invariant mass formed from two photons to compute
inter-calibration coefficients for the electromagnetic
calorimeter. The selection of $\pi^0$ candidates is based on two
isolated photon candidates where the invariant mass of the two
photons is close to the $\pi^0$ mass. The output of the
AlCaRaw trigger contains only the reconstructed hits forming the $\pi^0$ candidates. The typical event size is around
1.5~kB which corresponds to about one percent of the full event size.
\item ECAL $\phi$ symmetry: relies on the azimuthal ($\phi$) invariance of the energy
deposition in physics events. This trigger records all reconstructed
hits in the event. The typical event size is around 5~kB.
\item HCAL $\phi$ symmetry:  relies on the $\phi$ invariance of the energy
deposition in physics events. In order to extract low energy deposits, the hadron calorimeter is required to operate in
non zero-suppressed mode and the output contains mainly HCAL raw data. The
typical event size is large due to the non zero-suppressed HCAL output and is approximately 180~kB.
\item Isolated track: uses isolated tracks
to measure calorimeter response to single pions. The trigger requires an isolated track with a
large transverse momentum (at least 10\GeVc). Recent studies have shown
that this trigger can be included as an HLT trigger since the rate is
low ($\sim$10~Hz) even though the event size is rather large (hundreds
of kB).
\end{itemize}

For the CRAFT exercise, the AlCaRaw triggers were used with some relaxed criteria
in order to record events for testing and validating the workflow for
calibration.


\section{Monitoring of  the HLT performance during CRAFT }
\label{sec:performance}

During CRAFT many tools for monitoring the High-Level trigger were
developed and tested. In this section the monitoring of  the
performance of the muon High-Level triggers is presented, since they
were more relevant for the cosmic ray data.

\subsection{Online DQM}
\label{sec:OnlineDQM}

Trigger monitoring at CMS is performed at two different levels: the first makes use of monitoring software run directly in
the filter units (see Section~\ref{sec:HLTDQM}), the second consists of a quasi-online monitoring system which reads events after HLT
processing (see Section~\ref{sec:SMDQM}). The quasi-online monitoring is preferred,
since there are no limitations
on the allowed bandwidth, which is the case for  software running in the HLT.

\subsubsection{DQM in the High-Level Trigger}
\label{sec:HLTDQM}

In the HLT, the filter units process events at rates up to 100 kHz and produce a limited
number of histograms. The histograms are delivered from each filter unit to
the storage managers at the end of each luminosity section (corresponding to roughly 93 seconds of
data taking). Identical histograms are then summed together and sent
to a storage manager proxy server, which saves the histograms to files and serves them to DQM consumer applications
along with the events. The advantage of monitoring the data in the filter units is that the DQM consumers can access
all the events processed by the HLT, including events that will eventually be rejected at this stage. The main focus of the trigger DQM modules in the filter units is to monitor L1 and HLT scalers,
and compute rates and rejection factors for each trigger algorithm and filtering stage. During CRAFT fewer than ten histograms were created in this way. 

\subsubsection{Quasi-online DQM}
\label{sec:SMDQM}

In the HLT a dedicated DQM stream can be implemented that selects any type of events, based on HLT decisions.
During CRAFT the HLT typically did not reject any event and an unbiased fraction of events were randomly
selected to be sent to the DQM stream. This stream provides detector raw data and, on explicit request, additional HLT information. The DQM applications receive event data coming from the storage manager proxy server using the HTTP protocol.
These events can be delivered at a configurable rate, which was set to 25~Hz during CRAFT. The
quasi-online monitoring of the trigger includes monitoring of the L1 hardware
as well as monitoring of the level of
agreement between emulated data and hardware data. The emulator software reproduces the hardware decision
for a given trigger configuration, is used to compute trigger efficiencies offline, and allows comparisons with hardware data.

\subsubsection{Monitoring the performance of the Muon HLT}
\label{sec:muperf}

The muon HLT algorithms subdivide the HLT processing into two logical steps: Level-2 (L2) and Level-3 (L3). This nomenclature is common to all HLT triggers in CMS and echos a traditional three-level trigger system. L2 uses L1 muons as seeds and incorporates the full resolution of the muon system and the calorimeters. L3  includes the information from the tracker.
The muon trigger reconstruction detects muons by linking tracks in the muon system with deposits in the calorimeter consistent with minimum-ionizing particles and with tracks in the inner tracking system.

Muon HLT triggers used during CRAFT were those that were designed for use during LHC collisions.
Since cosmic ray muons only rarely cross the beamline, most of them will not satisfy these
muon HLT requirements. Moreover, the muon reconstruction algorithm used in the L3 step of the
HLT only performs the tracking within restricted regions around the L2 muon trajectory.  This restriction
shortens the time required for the L3 processing and allows these triggers to meet the timing constraints
of the HLT. This restricted regional tracking algorithm uses the beam spot position to determine the
tracking regions, which is not appropriate in the case of cosmic ray muons.
Even though not many events are expected to satisfy the muon triggers that require tracking,
those that do satisfy this requirement can be used to collect data for studying the performance of the muon HLT algorithms.

\begin{figure}[h]
  \begin{center}
    \includegraphics[width=0.48\linewidth]{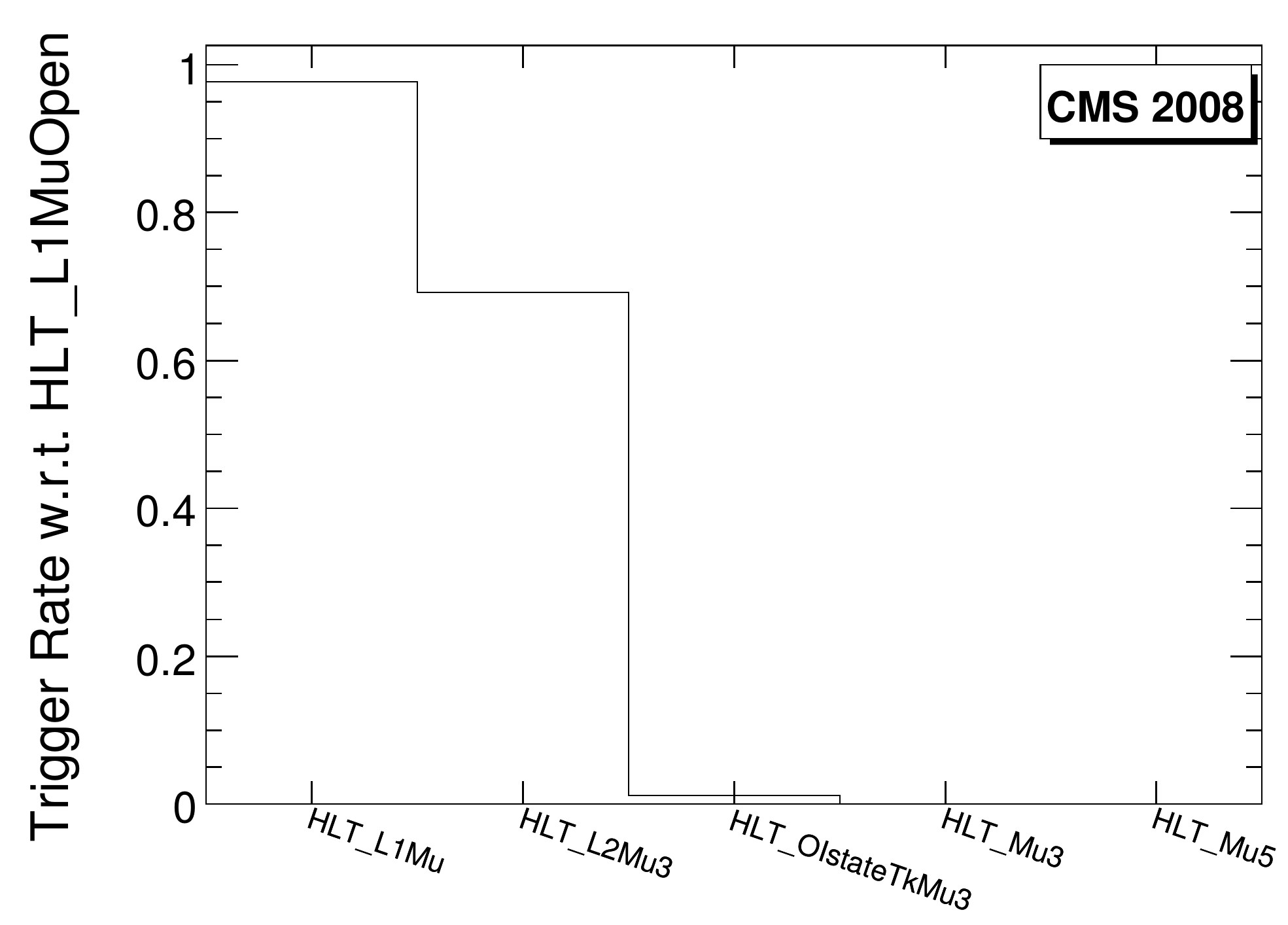}
    \includegraphics[width=0.48\linewidth]{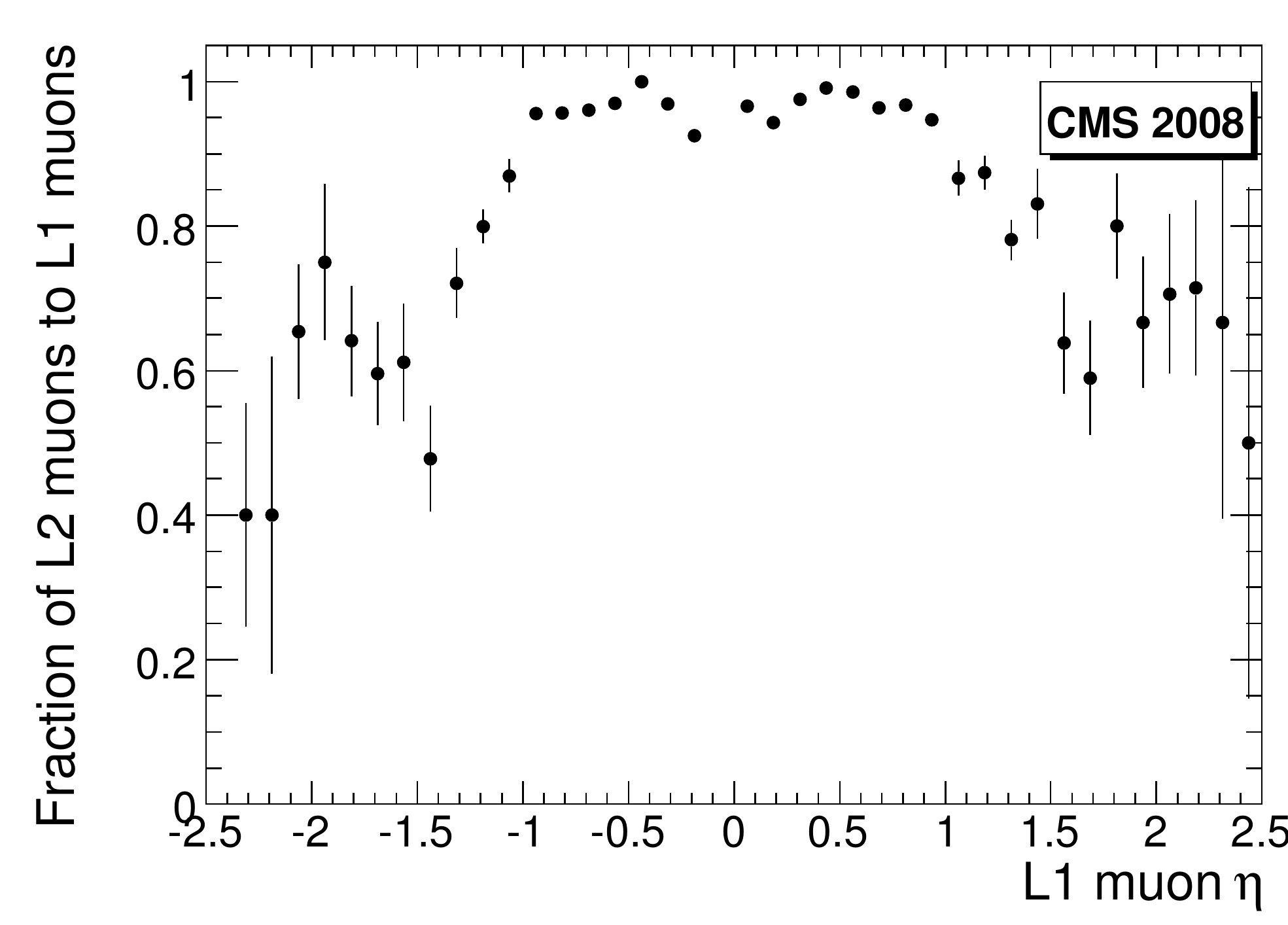}
    \caption{ Relative rate of the listed triggers with respect to the HLT\_L1MuOpen trigger (left) and the fraction of L2 muons to L1 muons as a function of the L1 muon pseudorapidity for events selected by the L2Mu3 trigger (right).
    }
    \label{fig:muonHLT}
  \end{center}
\end{figure}

The stability of the various triggers during collision data can be measured in terms of trigger rate and as a function of  the measured
luminosity. For cosmic data taken during CRAFT the stability was
monitored by comparing the rate of each trigger with that of the most
inclusive one, HLT\_L1MuOpen, a trigger requiring a L1 muon without any
momentum or quality restrictions. The performance of the latter was examined separately. The same scheme will be used with collision data and will provide a measurement of the trigger performance which does
not depend on uncertainties of luminosity measurements.
Figure~\ref{fig:muonHLT} (left) shows the rate for various muon triggers,
with respect to the HLT\_L1MuOpen trigger,
as measured in the online DQM during a good quality run of
50,000 events. This figure serves as a reference distribution and  is used to estimate the stability of HLT performance in other runs. The relative trigger rates depend on trigger and sub-detector configurations; therefore, a few reference distributions, one for each possible configuration, were prepared. The HLT trigger efficiency cannot be precisely measured in online DQM,
but the stability of trigger performance can be monitored by comparing
the number of L1 seeds and the number of L2 objects for a selected trigger as shown in Fig.~\ref{fig:muonHLT} (right). The figure represents the fraction of L2 muons with respect to the number of L1 muons, as a function of the L1 muon pseudorapidity $\eta$. Only events selected by a L2 trigger, L2Mu3, are considered. This figure serves as a reference distribution and changes to it indicate potential modifications in the performance of the muon trigger logic. A more detailed study of L2 and L3 muon trigger efficiencies and a comparison with offline reconstruction algorithms are presented elsewhere~\cite{CFT-09-014}.

\subsection{Offline low-latency high-statistics  data quality monitoring}

The offline reconstruction process at Tier-0 allocates computing
resources which can be deployed for systematic data quality
monitoring. ``Offline DQM'' software applications, similar to those
described for the online DQM, can be introduced as part of the offline
reconstruction task itself.  The primary disadvantage of monitoring
data in this fashion is the latency of offline reconstruction relative
to raw data collection, which can be as long as a few days.  The
complementary advantages of offline DQM are twofold: all reconstructed
events can be monitored, so that less obvious, slowly developing, or
rare problems can be identified; and the differences between online
and offline physics object reconstruction can be readily studied.
During CRAFT, offline DQM measuring of the HLT performance was exercised
as part of all offline reconstruction operations, and results were
available for study as part of the central DQM infrastructure for CMS.

\subsubsection{Software}

\begin{figure}[h]
   \begin{center}
      \includegraphics[width=0.52\linewidth,angle=90]{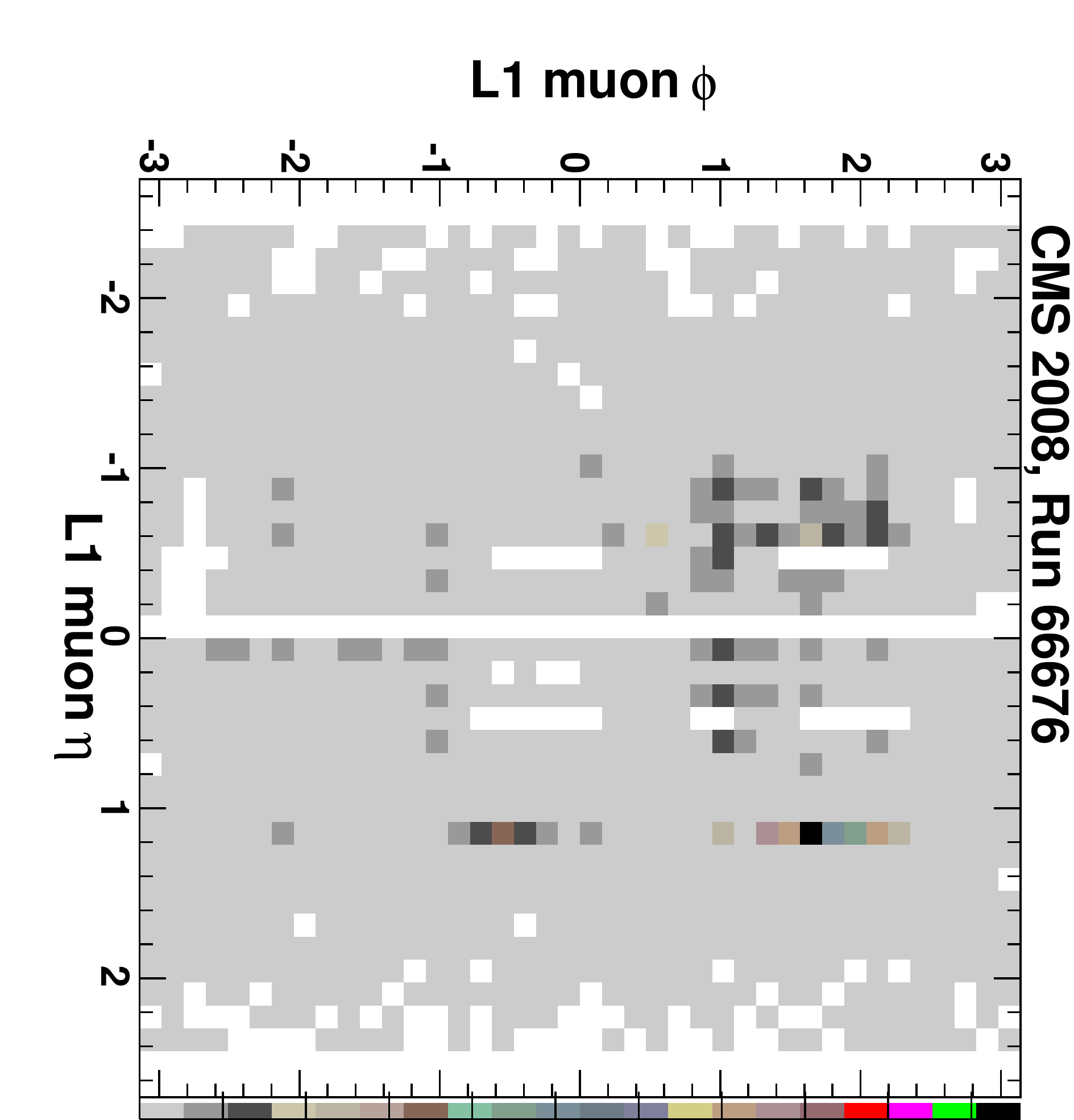}
      \caption{Example offline DQM occupancy plot for L1 muon candidates from a single CRAFT run, for events with high-quality cosmic muons reconstructed offline.}
   \label{fig:hltmuonoccupancy}
   \end{center}
\end{figure}

The software applications for offline DQM are identical in structure
to those for online DQM discussed previously.  The offline DQM produced the
following histograms: L1 trigger hardware performance plots, comparison of emulated L1 data with recorded L1 data, as described
in the previous section, and plots of basic properties of all objects
(muons, jets, etc., from L1, HLT, and offline reconstruction) passing
each of the HLT paths.  Plotted object properties include transverse
energy spectra, $\eta$-$\phi$ occupancy, and object multiplicity.  The
offline DQM software parses the HLT menu at run time and automatically
configures the correct set of plots for each trigger path in the menu.
Figure~\ref{fig:hltmuonoccupancy} shows the occupancy of muons which
passed a L1 muon path for a single CRAFT run.  With the large
sample size available offline, it is possible to identify regional
variations in muon trigger occupancy from run to run.

In addition to object properties recorded for each path, the offline
DQM software can require the objects to also pass a (configurable)
reference path, allowing for object sampling which is unbiased for
trigger efficiency studies.  For example, a minimum-bias reference
trigger path could be required, and offline muon candidates can be
plotted for all events in this reference path, and similarly for all
events which pass some muon trigger test path.  The quotient of the
transverse energy spectra of these two distributions is an estimator of the
test path trigger efficiency as a function of transverse energy for muons
reconstructed offline.  In preparation for proton collisions, more
specialized plots will be introduced.  Each trigger object type will
be analyzed separately, selecting high-purity and low-purity samples
for signal efficiency and background rejection studies.

\subsubsection{Usage for CRAFT}
The offline DQM applications ran regularly as part of offline
reconstruction throughout the CRAFT run, and the results were
available for viewing by offline DQM shifters.  Automated quality
tests and data certification were not yet deployed but are planned for
data taking.  The function of the offline DQM is largely to confirm
with higher precision the online DQM results.

Latency for offline DQM studies ranged from as little as an hour to as
long as a few days, depending on the reliability of the offline
reconstruction at any point in time.  No problems were originally
diagnosed offline, but it is anticipated that the increased HLT menu
complexity and longer integration times of proton collision runs will
require offline monitoring to discover more rare or more slowly
developing problems.


Other studies of the trigger conducted for the CRAFT run included:
emulation of the HLT algorithms offline and comparing the results with
the actual HLT data; studies of L1 and HLT performance of cosmic
muons; studies of the effect of calorimeter noise on L1 calorimeter
triggers, and observation of effects of noise and/or cosmic rays when a
menu anticipated for proton collisions was deployed.

\subsection{Monitoring of the CPU performance}

The average CPU time taken to accept or reject events at the HLT depends
significantly on the  type of event accepted by the L1 trigger. This
in turn depends on the sub-detectors included during the global data taking
process, the L1 and HLT prescales, and other operating
conditions. In order to track the behavior of the HLT CPU time under
distinct configurations, incremental CPU performance studies were carried out using
a few runs with different menus.  The results for two of these runs are presented in
this section.

Runs 66279 and 68021 were recorded during CRAFT using two different HLT
menus with different numbers of HLT paths. The sub-detector configuration
for these runs is summarized in Table~\ref{tab:runs}. In each case, the
performance of the full HLT menu was determined by measuring the total HLT
processing CPU time when running on a subset of the L1-accepted events. This total
CPU time (measured on a Dual-Core 5160 Xeon
processor running at 2.0~GHz) includes both the data-unpacking time and the time taken to execute the algorithms.The results scale by~0.7 since the machines used in the Filter Farm run on Quad-Core E5430 Xeon processors at 2.66~GHz.

\begin{table}[!Hhtb]
\begin{center}
\caption{Run configuration and total CPU time for runs 66279 and 68021
taken during CRAFT. \label{tab:runs}}
\begin{tabular}{|c|c|c|}
\hline
      & Run 66279                        & Run 68021\\ \hline \hline
Events used & 30000 & 30000 \\ \hline
 Sub-detectors & No ECAL, no tracker   & No HF \\
 included      & no RPC   & \\
  in run       &         &  \\ \hline
 Magnetic field & $0\ {\rm T}$  & $3.8\ {\rm T}$\\ \hline \hline
 \# of trigger paths &   16          &    24                 \\ \hline
 Average CPU time & 15~ms & 47~ms\\ \hline

\end{tabular}

\end{center}
\end{table}

The ECAL, the tracker,
and the RPC sub-detectors were not included in Run 66279.
Furthermore, the HLT menu used was rather simple, as it contained
only 16 triggers, which allowed primarily low quality muon objects to pass.

Run 68021 incorporated most of the detectors that were missing in the previous
run; however, the forward calorimeter (HF) was not included.
In addition, an increase in the number of triggers from $16$ to $24$
added complexity to the HLT menu.  The new
triggers included muon triggers with low thresholds that used tracking
information, a jet trigger, and a low threshold
trigger based on electromagnetic energy.
The complexity of the new algorithms (those
using tracking information in particular)
and the inclusion of the
missing sub-detectors are reflected in the three-fold increase
of the average CPU time.  Taking runs
with noisy sub-detectors allowed  gauging of the impact of non-optimal operating
conditions on the performance of the HLT. After a few such runs, only one algorithm was found to be non-optimal in terms of CPU performance. The data
collected during these special runs were used to develop more optimized algorithms that were subsequently deployed online. In one other case, corrupt data from one of the sub-detectors resulted in unexpectedly verbose error logging which in turn had an adverse effect on the the HLT processing time. Since this algorithm had previously been extensively and successfully tested on simulated events, this incident reiterated the importance of commissioning the HLT algorithms with data. Following the CRAFT data taking period all HLT algorithms were scrutinized in detail to ensure that error logging was kept to a minimal and that corrupt data were instead detected via DQM applications.

The unpacking of L1 data took 3 to 4~ms and the remaining CPU time was taken
up by sub-detector unpacking (based on L1 decisions)
and the execution of the HLT algorithms.
The L1 unpacking time was approximately the same for all the
studies described in this section.  This is the expected behavior since the
addition of new triggers to a menu does not increase the CPU time for the
L1 data unpacking, as this is carried out only once per event.

This monitoring of  the CPU performance of the HLT allowed testing of the incremental stability of the system under pronounced variations of conditions and sub-detector configurations.



\section{Summary and Conclusions}
\label{sec:conclusion}
The commissioning efforts for the High-Level Trigger in 2009 have built on the
progress made during the 2008 CRAFT period.  During CRAFT, physics-based selection criteria beyond the L1
decision were integrated into the HLT menus and the latter were run successfully  during an extended period of time.  The ability to reliably and efficiently collect physics-enriched datasets is critical for collisions, and CRAFT proved invaluable for commissioning HLT operations.  We monitored the overall performance of the HLT and confirmed its dependence on trigger design: triggers that are not robust or are not optimized for online performance
greatly affect the stability of HLT operations.

These lessons have significantly influenced recent commissioning work.  The HLT
menus for collisions have been redesigned in preparation for data taking with greater emphasis being paid to simple, inclusive menus with a small set of robust triggers.   The triggers present in these new, lean menus are being scrutinized to maximize trigger performance; this includes optimizing the choice of L1 seeds for each trigger, so as to select those events considered most likely to pass the HLT selection algorithms.  Reconstruction algorithms are also being analyzed for performance to ensure that fast, reliable trigger decisions can be reached regardless of detector conditions.

Although individual triggers have been extensively tested on simulated events, changes in data quality delivered by the CMS detector (e.g., noisy
or corrupt data) can dramatically influence the trigger behavior.  For this
reason, HLT DQM operations serve as an important check of the trigger
performance.  Any change in the trigger behavior can be quickly detected and prompt experts to implement fixes to the existing triggers or design more
robust algorithms for future menus. HLT DQM has improved considerably since 2008 as a direct result of the CRAFT experience.

The alignment and calibration workflows were successfully validated during CRAFT. The AlCaRaw datasets were rapidly made available and were used to study detector performance.  This demonstrated that the HLT can efficiently provide large datasets that can be used for a rapid calibration of the CMS detector during collision data taking.  These efforts have continued in 2009 and remain a successful part of the HLT commissioning process.

\section*{Acknowledgements}

We thank the technical and administrative staff at CERN and other CMS Institutes, and acknowledge support from: FMSR (Austria); FNRS and FWO (Belgium); CNPq, CAPES, FAPERJ, and FAPESP (Brazil); MES (Bulgaria); CERN; CAS, MoST, and NSFC (China); COLCIENCIAS (Colombia); MSES (Croatia); RPF (Cyprus); Academy of Sciences and NICPB (Estonia); Academy of Finland, ME, and HIP (Finland); CEA and CNRS/IN2P3 (France); BMBF, DFG, and HGF (Germany); GSRT (Greece); OTKA and NKTH (Hungary); DAE and DST (India); IPM (Iran); SFI (Ireland); INFN (Italy); NRF (Korea); LAS (Lithuania); CINVESTAV, CONACYT, SEP, and UASLP-FAI (Mexico); PAEC (Pakistan); SCSR (Poland); FCT (Portugal); JINR (Armenia, Belarus, Georgia, Ukraine, Uzbekistan); MST and MAE (Russia); MSTDS (Serbia); MICINN and CPAN (Spain); Swiss Funding Agencies (Switzerland); NSC (Taipei); TUBITAK and TAEK (Turkey); STFC (United Kingdom); DOE and NSF (USA). Individuals have received support from the Marie-Curie IEF program (European Union); the Leventis Foundation; the A. P. Sloan Foundation; and the Alexander von Humboldt Foundation.

\bibliography{auto_generated}   
\cleardoublepage\appendix\section{The CMS Collaboration \label{app:collab}}\begin{sloppypar}\hyphenpenalty=500\textbf{Yerevan Physics Institute,  Yerevan,  Armenia}\\*[0pt]
S.~Chatrchyan, V.~Khachatryan, A.M.~Sirunyan
\vskip\cmsinstskip
\textbf{Institut f\"{u}r Hochenergiephysik der OeAW,  Wien,  Austria}\\*[0pt]
W.~Adam, B.~Arnold, H.~Bergauer, T.~Bergauer, M.~Dragicevic, M.~Eichberger, J.~Er\"{o}
% ...
\vskip\cmsinstskip
% ...

\end{sloppypar}
\end{document}